\documentclass{article}
\input amssym.def
\input amssym.tex
\def\ben{\begin{equation}}
\def\een{\end{equation}}
\def\half{{1/2}}
\def\bea{\begin{eqnarray}}
\def\eea{\end{eqnarray}}

\begin{document}

{\obeylines

           COSMIC BALDNESS 

           W. Boucher 
           G.W. Gibbons 
           D.A.M.T,P. , 
           Cambridge university 

}

\vskip 2cm 
\medskip 
{\underline{INTRODUCTION}}
\medskip
 
Some years ago (Gibbons \& Hawking 1977) it was suggested that solutions of Einstein's equations with a positive cosmological term should eventually settle down to a state which is stationary inside the cosmological event horizon of any future inextendible timelike curve. This stationary state, it was suggested, would, if no black holes were present, be de Sitter space. The recent work on the inflationary scenario in the early universe has reawakened interest in this topic and the purpose of this article is to review the mechanism whereby a universe dominated by vacuum energy, so that it satisfies the equation 

\ben
R_{\alpha \beta }= \Lambda g_{\alpha \beta}\,, \qquad  \Lambda >0\,,
\een
relaxes to an asymptotically de sitter state inside the event horizon of any observer. There is some overlap with the article by Barrow but the emphasis here is rather different. 

\medskip
{\underline {GEOMETRY OF DE  SITTER  SPACE}}
\medskip 

De Sitter space may be thought of as the hyperboloid in 5 dimensional Minkowski space given by 
\ben
-(X^0)^2 +(X^1)^2+ (X^2)^2+ (X^3)^2 + (X^5)^2 = 3/ \Lambda 
\een
where $\Lambda $ is the cosmological constant which is related to Hubble's
constant, $ H$ , or the surface gravity of the horizon, $\kappa$,  by 
\ben
H= \kappa = ({\Lambda \over 3} ) ^ {\half}     
\een

It is useful to coordinatize the spacetime in two different ways,
depending upon whether we wish to think of it as an expanding F.R.W. 
universe or a static universe with an event horizon. We use 
$(s,\chi, \theta,\phi)$  
the first case and $(t,r,\theta ,\phi)$  in the second. They are given (Hawking 
\& Ellis 1973; Hawking \& Gibbons 1977): 

\bea
r\sin \theta \cos \phi&=& X^1 = H^{-1} \cosh Hs \,\sin \chi \sin \theta
\cos \phi\\
r\sin \theta \sin  \phi&=& X^2 = H^{-1} \cosh Hs\, \sin \chi \sin \theta
\sin \phi\\
r \cos \theta &=&X^3 = H^{-1} \cosh Hs \,\sin \chi \cos  \theta \\
( H^{-2} -r^2 ) ^\half \cos Ht &=& X^5 = H^{-1} \cosh Hs \cos  \chi \\  
( H^{-2} -r^2 ) ^\half \sin  Ht &=& X^0 = H^{-1} \sinh Hs  
\eea

The metric thus becomes: 
\ben
d {\frak{ s}} ^2= -(1- H^2 r^2 ) dt ^2 + (1-H^2 r^2 ) ^{-1} dr ^2 + r^2
d\Omega ^2  
\een
or 
\ben
d {\frak{ s} }^2 = 
-ds^2  + H^{-2}  \cosh ^2 H s (d \chi ^2  + \sin^2 \chi d \Omega ^2  ) 
\een
The event horizon of an observer situated at $r=0$ 
given by $r =H^{-1}$ and so since (4) - (6) imply that 
\bea
\sin \chi &=& Hr (\cosh Hs)^{-1}\\&=& Hr (1-H^2 r^2 \tanh^2 Ht) ^{-\half}
(\cosh Ht ) ^{-1} 
\eea
we see that as $s,t \rightarrow$  an exponentially 
smaller portion of the 3-spheres $s =$  const. is 
included within the event horizon of this observer. 
This is the key to understanding the decay of perturbations. 
From the point of view of the static frame we expect them to be 
radiated through the cosmological event horizon just as in the 
familiar black hole case (see Price 1972). 
From the point of view of the expanding universe 
frame we shall see --  as pointed out by Starobinsky (1977) 
that gravitational wave perturbations do not decay but rather 
they are frozen  in as $t\rightarrow \infty$ . 
However this is not in contradiction with the cosmic No 
Hair Theorem since as time proceeds the observer sees these perturbations on an exponentially smaller scale and so the region inside his/her event horizon appears to become more and more accurately de Sitter. 

\medskip
{\underline { BEHAVIOUR  OF LlNEARIZED  SCALAR AND GRAVITATIONAL WAVE}}
{\underline { 
PERTURBATIONS}  }\medskip

Following Lifshitz \& Khalatnikov (1963) we consider perturbations of
the metric form (1O) given by
\ben
d {\frak{ s}}  ^2 =-ds^2 + H^{-2} \cosh ^2 Hs (d \chi ^2 + \sinh ^2 \chi d
\Omega ^2 + \sum _n \nu _n(s)  G^{(n)}_{ij} (\chi,\theta \phi) dx ^i dx ^j
) 
\een 
where the $G^{n}_{ij}$ are tensor harmonics on $S^3$ and the $\nu_n(s)$ are
the amplitudes of the gravitational wave perturbations.
It turns 
are the out that if one considers solutions of the massless, 
minimally coupled scalar wave equation in the de Sitter background of the form 
\ben
 \phi = \nu  _n(s) Q^{(n)} (\chi,\theta ,\phi)  
\een
where the $Q^{n} $  are scalar harmonics on $S^3$ , 
the coefficients $\nu_n(s)$ in (13) satisfy the same equation as those in (12) 
and are given constant factor by 
\ben
\nu_n(s) = 
(in \,  {\rm sech}  Hs + \tanh Hs) \exp in( \tan^{-1} \sinh Hs )  \sinh HS) 
\een  
In neither the scalar nor the gravitational case do the 
perturbations die away, rather they tend to constants at late times. 
Thus the scalar field $\phi$   can have any functional form,
$\Phi(\chi,\theta ,\phi) $, at late times. 
However using the relation given by (l1) between $\chi$ 
 and the coordinates $r$ and $t$  we see that as $ t \rightarrow
 \infty $ for all $r$  inside the 
event horizon 
\ben
\Phi(t,r, \theta,\phi) \rightarrow \Phi(0) 
\een
Thus $\phi$  tends to a constant inside the event horizon exponentially fast which is in accordance with the fact that there are no static solutions of the wave equation which are regular inside and on the event horizon 
other than the constant one. 

In the gravitational case similar results hold, locally the 
constant gravitational wave modes are pure gauge, even though they are 
not pure gauge globally over the entire $S^3$ .
We shall see this in more detail when we consider the fully
non-linear asymptotic form of the metric. Note that in contrast to
the black hole case the perturbations die away exponentially fast;
there is no power law fall-off 
of the sort discussed by Price (1972). 

\medskip 
{\underline{ FULLY NON-LINEAR ASYMPTOTIC ANALYSIS }}
\medskip

Starobinsky has pointed out to us that a general asymptotic 
solution of the equation $R_{\alpha \beta}= \Lambda g_{\alpha \beta}$  
takes the form 
\ben
d{\frak{s}}^2 = - ds ^2 + \exp {2Hs} \,a_{ij} ({\underline x}) dx^i dx^j
+ O(1) \een 
where $a_{ij}$ is  
is an $\underline{\rm arbitrary}$  3-metric. 
This clearly indicates that the 
waves do not decay globally over the entire 3-surface 
s = constant. The geometry af this surface never settles down to 
that of a smooth 3-sphere. The curve $x^i = 0$ 
 is a geodesic. By means of a linear 
coordinate transformation of the $x^i$ 's we may, with no loss of
generality, set 
\ben
a_{ij}(0)= \delta _{ij}
\een
Now introduce coordinates $y^i$ and $t$ by  
\bea
y^i&=& e^{Hs} x^i\\
e^{Ht}&=& (1-H ^2 y^2) ^{-\half} e^{Hs}
\eea  
It is now an easy exercise to show that in the coordinates $(t, y^i)$, 
which are only valid within the event horizon of the timelike observer
at 
$y^i =0$ , the general metric (16) approaches the exact metric of 
de Sitter space (equation (9))  exponentially fast. 
Thus as far as every freely falling observer is concerned the
observable 
universe becomes quite bald.

\medskip 
{\underline { GENERALIZED  ISRAEL' S THEOREM} } 
\medskip
 
If we exclude the possibility of a central black hole it 
would seem likely from the above analysis that de Sitter space is the
only exactly static solution of the equations $R_{alpha \beta}=\Lambda
g_{\alpha \beta}\,, \quad \Lambda >0$, surrounded by a regular 
event horizon and with a regular centre -- i.e. such that the interior
of the event horizon is diffeomorphic to the product of an open ball
in $R^3$
 with the real line. This last proviso is necessary to exclude the 
Nariai (1951) metric. We have tried unsuccessfully to generalize the standard Israel theorems for black holes (Israel 1967; Muller zum Hagen, et al. 1973; Robinsan 1977) to this case. If the 
metric is static the boundary conditions on the horizon are just those required to justify analytically continuing the metric to the Euclidean 
regime by putting $t = i\tau$. 
The resulting
metric is positive definite, defined on a manifold diffeomorphic to
$S^4$, and satisfies $R_{\alpha \beta}=\Lambda g_{\alpha \beta }$. 
 standard Einstein metric on 
The only known such metric is of course the standard 
Einstein metric on $S^4$  
which is just the analytic continuation of 
de Sitter space obtained   by setting 
$t=i\tau$ in (9) amd identifying $\tau$  , 
modulo $2\pi ( 3/\Lambda) ^\half$ . 
The natural conjecture to make -- which is stronger than the
generalised  Israel theorem suggested above -- is that the only such
metric is the standard one. Mathematicians we have asked
do not know whether this is true  but the corresponding statement 
is false for $S^{4n+3}\,,\quad n\ge 1$
(Jensen 1973). If there is another such metric it cannot 
be continuously deformed into the standard one as one can readily
check 
by perturbing 
the Einstein equations on $S^4$  as described in (Gibbons and Perry 1978). 

\medskip
{\underline{ ACKNOWLEDGEMENTS}}
\medskip

We would like to thank S.W. Hawking, A. Starobinsky, 
S. Siklos and J. Barrow for useful discussions and suggestions on the material presented above.

\medskip
{\underline{ REFERENCES}} \medskip

\medskip \noindent 
Gibbons, G.W., and Hawking, S.W. (1977). cosmological event horizons,
            thermodynamics, and particle creation. Phys. Rev. D { $
            \underline{15}$}, 2738. 

\medskip \noindent
Gibbons, G.W., and Perry, M.J. (1978). Quantizing gravitational-instantons. 
               Nuc1. Phys. B {$ \underline  {146} $} , 90. 

\medskip \noindent
Hawking, S.W., and Ellis, G.F.R. (l973)  The large scale structure of 
             space-time. Cambridge: Cambridge University Press. 

\medskip \noindent
Israel, W. (1967)  Event horizons in static vacuum space-times. Phys. 
                   Rev. {$\underline {164} $} , 1776. 

\medskip \noindent
Jensen, G.R. (1973). Einstein metrics on principal fibre bundles.

                     J. Diff. Geom.{$\underline  8$}  599  

\medskip \noindent
Lifshitz, E.M. and Khalatnikov, I.M. (1963). Investigations in

                   Relativistic Cosmology. Adv. Phys. {$ \underline {12} $}, 185. 

\medskip \noindent
Muller zum Hagen, H., Robinson, D.C., and Seifert; H.J. (1973). Black 
                   holes in static vacuum space-times. 
                   Gen. Rel. Grav.{$\underline 4$} 53.

\medskip \noindent
Nariai, H. (1951). On a new cosmological solution of Einstein's field 
                   equations of gravitation. Sci. Rep. Tohoku
                   Univ. {$\underline  {35} $} , 62. 

\medskip \noindent
Price, R.H. (1972). Nonspherical perturbations of relativistic 
                    gravitational collapse, I: Scalar and gravitational
                    perturbations. Phys. Rev. D{$\underline 5 $ },
                    2419. 
                    Nonspherical perturbations of relativistic
                    -gravitational collapse, II: 
                    Integer-spin, zero-rest-mass fields. Phys. Rev.
                    D{$\underline 5$} , 2439. 

\medskip \noindent
Robinson, D.C. A simple proof of the Generalization of Israel's Theorem. 
                    Gen. Rel. Grav. {$\underline 8$} , 695. 

\medskip \noindent
Starobinsky, A.A. Spectrum of relict gravitational radiation and the 
                   early state of the universe. JETP
                   Lett. {$\underline {30} $} , 682

\end{document}